\documentclass[12pt]{article}
\usepackage{epsfig}

\newcommand{\anti}[1] {$\mathrm{\overline {#1}}$}
\newcommand{\be}{\begin{equation}}
\newcommand{\ee}{\end{equation}}
\newcommand{\Journal}[4]{{#1} {\bf #2}, #3 (#4)}
\newcommand{\ZPC}{{\em Z. Phys.\ }C}

\begin{document}
\begin{titlepage}
\begin{flushright}
NORDITA-99/73 HE
\end{flushright}

\begin{center}
{\bf Event Selection Effects on Multiplicities in Quark and Gluon Jets}
\vspace{2cm}

P. Ed\'en

{\it NORDITA, Blegdamsvej 17, DK-2100 Copenhagen, Denmark\\E-mail: eden@nordita.com}

\vspace{3cm}
\begin{quotation}
\noindent
The properties of quark and gluon jets depend on jet definitions and event selection. I discuss how these can be included in calculations and present jet definitions designed to give unbiased jets.
\end{quotation}

\end{center}
\end{titlepage}

\section{Introduction}\label{sec:intro}
Perturbative QCD predicts the ratio of multiplicities in quark and gluon jets to reach an asymptotic value $C_{\mathrm{A}}/C_{\mathrm{F}}=9/4$. At accessible energies, it is however found to be substantially lower, $\sim 1.5$ at $90\mathrm{GeV}$.\cite{OPAL} Instead of a method to establish the colour factor ratio, the average multiplicities are a simple and powerful probe of subleading corrections. Energy conservation effects on the multiplicity ratio have been studied\cite{Dremin+Ochs,qgjets} and results are in agreement with data.\cite{DELPHI}

As there is no colourless point source for gg events available at energies above onium resonances, the major way to study scale evolution of $N_{\mathrm{gg}}$ is via jets. However, event selection and jet definitions introduce a bias that needs to be well understood. That is the issue of this talk, with focus on $e^+e^-$ annihilation.

\section{Two-jet Events}\label{sec:2jet}
\begin{figure}[t]
\parbox{0.57\textwidth}{
\begin{center}
\epsfxsize=0.35\textwidth
\epsfbox{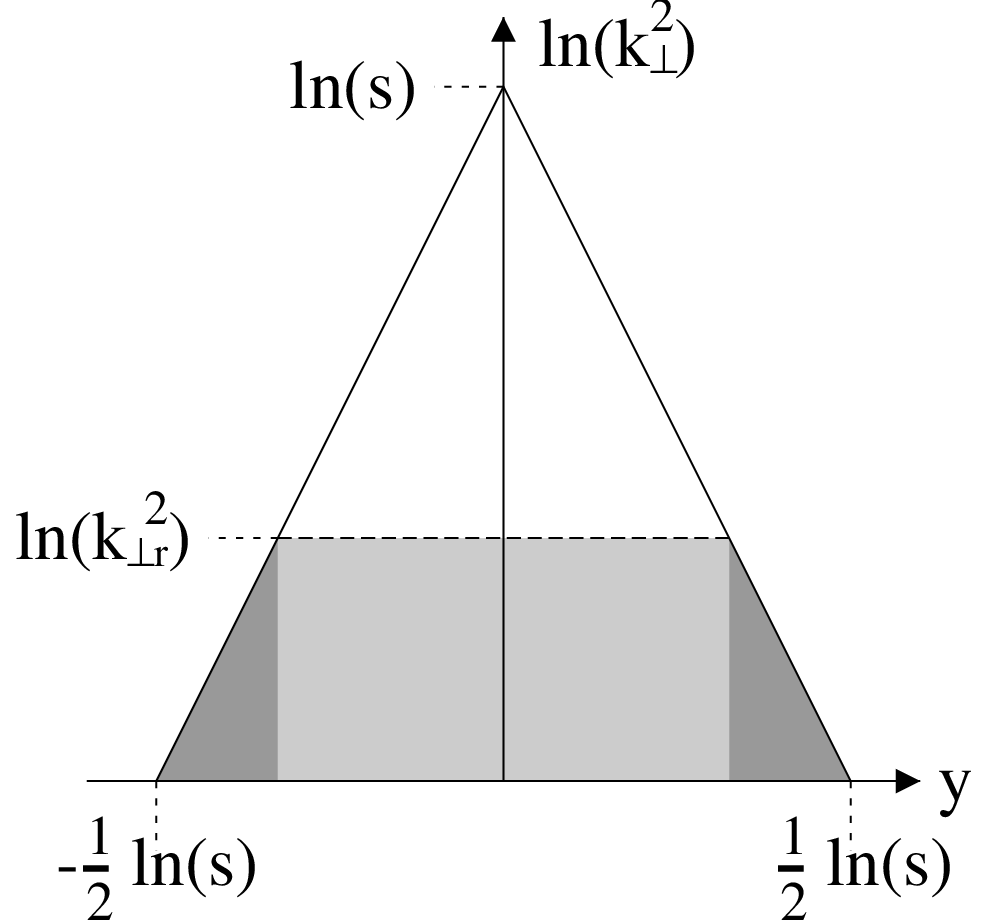}
\end{center}
\caption{
In a two-jet event sample selected with a resolution scale $k_{\perp r}$, gluon momenta are constrained by  $k_\perp < k_{\perp r}$, in addition to the kinematical limit $|k|<\sqrt s/2$ which corresponds approximately to the triangle $|y|<\frac12\ln(s/k_{\perp}^2)$
. The multiplicity in the forward cones of the jets (dark gray) is unbiased, while the multiplicity in the central region (light gray) is constrained by $k_{\perp r}$.\label{f:ktr}} 
}
\hspace{0.04\textwidth}
\parbox{0.37\textwidth}{
\epsfxsize=0.36\textwidth
\epsfbox{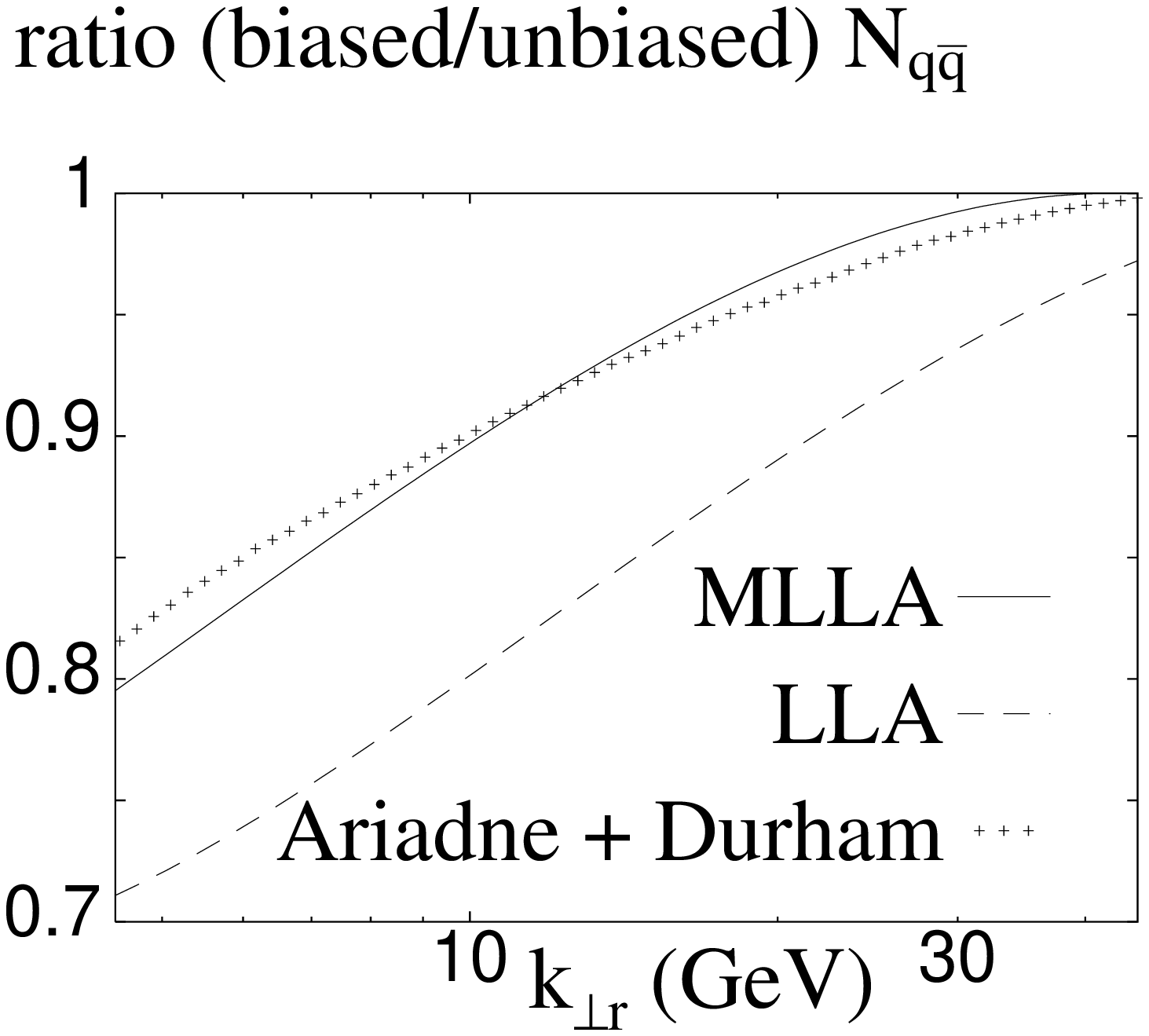}
\caption{The ratio biased over unbiased multiplicity. The bias gives a correction larger than 10\% for $k_{\perp r} <20\mathrm{GeV}$. The MLLA prediction is in agreement with MC simulations using the Durham jet finder. 
\label{f:Nqqbias}}
}
\end{figure}
We start to investigate the bias in the simple case of two-jet events selected with a jet clustering algorithm with a resolution scale $k_{\perp r}$. The situation is illustrated in Fig.~(\ref{f:ktr}), in the plane of logarithmic variables $\ln(k_\perp^2/\Lambda^2)$ and rapidity $y$. This coordinate system is well suited to illustrate the multiplicity formulae below, as the emission density of gluons is essentially flat in rapidity.

In a central rapidity range the upper constraint on subjet transverse momenta is $k_{\perp r}$, but in a forward cone of each jet the kinematical constraint is more restrictive.
Summing up the multiplicities in the forward cones and the central region, we get\cite{earlybias,qgjets}
\be
N_{\mathrm{q\overline q}}(s,k_{\perp r}^2) = N_{\mathrm{q\overline q}}(\zeta k_{\perp r}^2)+\ln\left(\frac{s}{\zeta k_{\perp r}^2}\right)\left.\frac{\partial N_{\mathrm{q\overline q}}(Q^2)}{\partial\ln Q^2}\right|_{Q^2=\zeta k_{\perp r}^2},~~\zeta=  e^{\frac32}. \label{e:Nqqbias}
\ee
The scale shift factor $\zeta$ is a modified leading log (MLLA) correction to the leading order result $\zeta=1$.\cite{qgjets}

Fig.~(\ref{f:Nqqbias}) shows the ratio biased over unbiased multiplicities for different resolution scales $k_{\perp r}$ at fixed energy  $\sqrt s=90\mathrm{GeV}$. To calculate the biased multiplicity in Eq.~(\ref{e:Nqqbias}) we have used a simple fit to \textsc{Ariadne}+\textsc{Jetset} MC\cite{MC} results for the unbiased $N_{\mathrm{q\overline q}}$.
As seen, the effect of the MLLA correction factor $\zeta$ is significant. The bias is in general important, being more than a 10\% effect for $k_{\perp r}<20\mathrm{GeV}$.
Also shown in  Fig.~(\ref{f:Nqqbias}) are MC results with \textsc{Ariadne} and the Durham jet algorithm.\cite{durham}

\section{Three-jet Events}\label{sec:threejet}
\begin{figure}[t]
\begin{center}
\epsfxsize=0.98\textwidth
\epsfbox{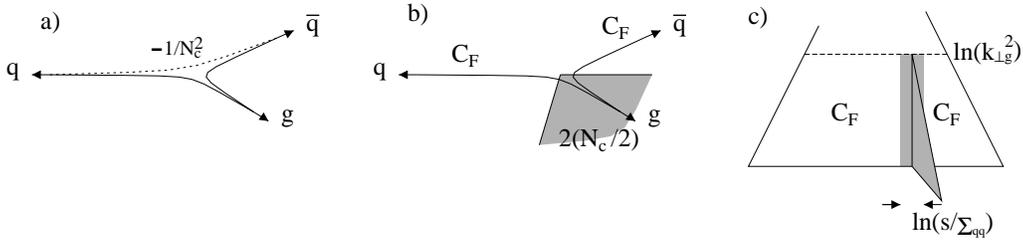}
\end{center}
\caption{The q\anti qg configuration. {\bf (a)} The emission density corresponds to two dipoles plus a colour suppressed correction here represented by a dipole spanned between the q and \anti q. {\bf (b)} The correction term can be approximated by a colour factor reduction $N_{\mathrm{c}}/2\rightarrow C_{\mathrm{F}}$ in the quark ends of the two dipoles attached to the gluon. {\bf (c)} In the logarithmic phase space variables, the two dipoles can be represented by the original q\anti q triangle plus a double sided fold. The magnitude of the $C_{\mathrm{F}}$ regions in the q and \anti q jets introduces a new scale, here called $\Sigma_{\mathrm{q\overline{q}}}$. 
\label{f:qqg}}
\end{figure}
The multiplicity in gluon jets can be extracted from q\anti qg three-jet events.
If a three-jet event sample is selected with a fixed resolution $k_{\perp r}$, we get complicated scale dependences, and all jets are biased. It is therefore more suitable to perform iterative clustering until exactly three jets remain, and study the event as a function of the $k_{\perp \mathrm{g}}$ of the softest jet, presumably the gluon jet. In this approach the gluon jet is essentially unbiased, but the bias in the q and \anti q jets needs still to be considered.

The q\anti qg configuration is illustrated in Fig.~(\ref{f:qqg}).
The emission density of softer gluons corresponds to radiation from a qg- and a g\anti q colour dipole.\cite{leningrad?}
There is also a colour correction term, corresponding to a q\anti q dipole, whose contribution is weighted with the negative factor $-N_{\mathrm c}^{-2}$.
As apparent from the negative weight, this colour correction dipole is not an independent emitter. Instead it can be assumed to reduce the colour factor from $N_{\mathrm{c}}/2$ to $C_{\mathrm{F}}$ in the quark ends of the two other dipoles.
The magnitude of the $C_{\mathrm{F}}$ phase space regions can not be determined by perturbative QCD, since the notion of an  infrared cut-off independently defined for each dipole does not apply to the colour correction term. This introduces an uncertainty of relative order $\frac 1{N^2_{\mathrm c}}\frac1{\ln s}$ to the total hadronic multiplicity\cite{qgjets} and hinders the possibility to fully determine corrections of relative order $1/\ln s$ from first principles. 

\begin{figure}[t]
\begin{center}
\epsfxsize=0.5\textwidth
\epsfbox{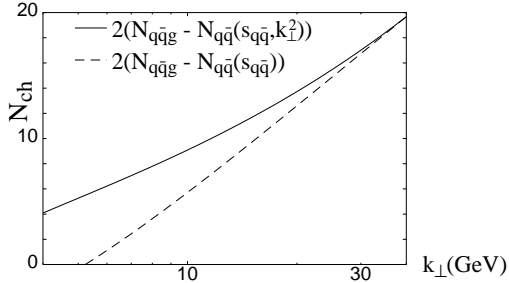}
\end{center}
\caption{$N_{\mathrm{gg}}$ extracted from three-jet events with $\sqrt s=90\mathrm{GeV}$ and $\sqrt{\Sigma_{\mathrm{q\overline{q}}}}=60\mathrm{GeV}$. Neglecting the bias (dashed line) underestimates $N_{\mathrm{gg}}$ at low scales, as compared to the result when the bias is taken into account (solid line). \label{f:ngg}}
\end{figure}

The magnitude of the $C_{\mathrm{F}}$ regions in the q and \anti q jets introduces a new scale, here called $\Sigma_{\mathrm{q\overline{q}}}$. 
Provided $\Sigma_{\mathrm{q\overline{q}}}$ is not much smaller than $s$, the total multiplicity for the three-jet event is\cite{letter}
\be
N_{\mathrm{q\overline{q}g}}  \approx  N_{\mathrm{q\overline q}}(\Sigma_{\mathrm{q\overline{q}}},k_{\perp \mathrm{g}}^2)+\frac12N_{\mathrm{gg}}(k_{\perp \mathrm{g}}^2\frac s{\Sigma_{\mathrm{q\overline{q}}}}) \label{e:nqqg}.
\ee
This equation shows the possibility to extract unbiased $N_{\mathrm{gg}}$ using three-jet event data. The importance of taking the bias on the q\anti q system into account is illustrated in Fig.~(\ref{f:ngg}). The result is sensitive also to the assumed value of $\zeta$ in Eq.~(\ref{e:Nqqbias}). Neglecting the MLLA corrections to the bias, by setting $\zeta=1$ in Eq.~(\ref{e:Nqqbias}), raises the prediction on $N_{\mathrm{gg}}$ by about two to three charged particles. This  emphasizes the importance of confronting Eq.~(\ref{e:Nqqbias}) with data.

Though the dependence on $\Sigma_{\mathrm{q\overline{q}}}$ is formally suppressed by $1/N_{\mathrm{c}}^2$, different $\Sigma_{\mathrm{q\overline{q}}}$ values implicitly correspond also to different recoil assumptions.
The recoil treatment when q\anti qg $\rightarrow$ q\anti qgg can in principle be constrained by comparisons with the four parton matrix element,\cite{4match} but an exact solution is beyond reach within a parton cascade formalism. This puts  a limit on the accuracy in the analysis presented here.
The prediction on $N_{\mathrm{q\overline{q}g}}$ for the reasonable assumptions $\Sigma_{\mathrm{q\overline{q}}} = s_{\mathrm{q \overline q}}$ and $\Sigma_{\mathrm{q\overline{q}}} = s$ 
differ by about $\frac12$ charged particle when $\sqrt s = 90\mathrm{GeV}$ and $\sqrt {s_{\mathrm{q\overline{q}}}}=60\mathrm{GeV}$.\cite{letter}

\section{Individual Jets}\label{sec:jets}
The method described in the preceding section has the advantage that we need no explicit investigation of individual jets, but it may suffer from dependences on recoil assumptions. In this section we discuss the scale dependences in jets, and present a jet definition designed to give unbiased gluon jets. This gives an alternative way to observe the scale evolution of the unbiased $N_{\mathrm{gg}}$.

The properties of a jet depend on (at least) two scales. The maximum $k_\perp$ for subjets and the available rapidity range $Y_j$ (at some cut-off scale $\Lambda$) in the jet, which depends on the jet energy. The two scales are truly independent, and can not be combined into one single effective scale. As an example, each hemisphere of a two-jet event illustrated in Fig.~(\ref{f:ktr}) can be seen as a quark jet with transverse momentum scale $k_{\perp r}$ and a rapidity scale $\frac12\ln(s/\Lambda^2)$.

In general, $Y_j$ is not easily determined. Each jet can get multiplicity contributions from several dipoles, these dipoles are boosted away from their rest frame, and the opening angle of the jet need not be azimuthally symmetric. 
\begin{figure}[t]
\parbox{0.36\textwidth}{
\epsfxsize=0.32\textwidth
\epsfbox{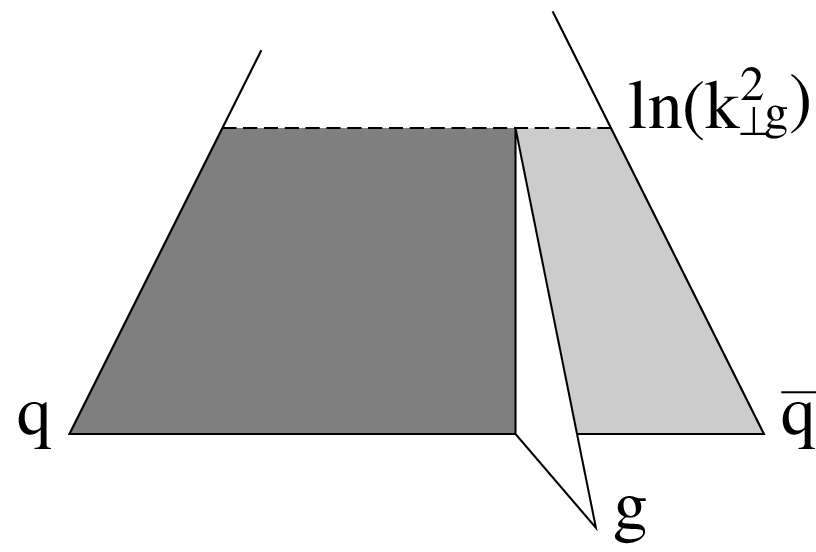}
\caption{The ``Boost'' jet boundaries. The gluon jet is unbiased and the quark jets have well defined rapidity scales $Y_{\mathrm q}$. \label{f:BoostCE}}
}
\hspace{0.05\textwidth}
\parbox{0.57\textwidth}{
\begin{center}
\epsfxsize=0.44\textwidth
\epsfbox{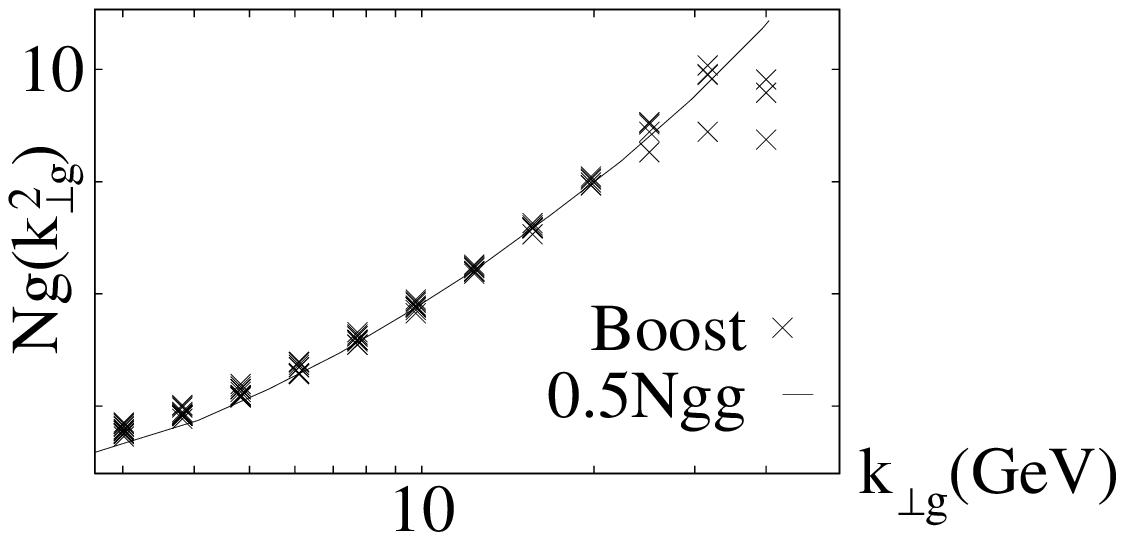}
\end{center}
\caption{MC simulations with the Boost method. For fixed $k_{\perp \mathrm{g}}$ the charged multiplicity in the gluon jet is essentially independent of $y$ and in good agreement with MC simulations of $\frac12N_{\mathrm{gg}}$.  \label{f:Ng}}
}
\end{figure}
In order to get better control of the $Y_j$ scale, we suggest a jet analysis in two steps
\begin{itemize}
\item Find jet {\em directions} using a $k_\perp$-based cluster algorithm.
\item Redefine jet {\em boundaries} to get well controlled $Y_j$ values.
\end{itemize}
With this approach it is possible to define jet regions as illustrated in Fig.~(\ref{f:BoostCE}).\cite{qgjets} This implies that we can study unbiased gluon jets and also quark jets with well defined rapidity scales
\be
Y_{\mathrm{ q}(\mathrm{\overline{q}})} = \frac12\ln(s)\, +\!(\!-\!)\, y \label{e:Yqboost},
\ee
where $y= \frac12\ln[(1-x_{\mathrm{\overline{q}}})/(1-x_{\mathrm q})]$ is the rapidity of the gluon jet.

The boundaries are defined in terms of a Lorentz boost which is described in detail elsewhere.\cite{qgjets}
Fig.~(\ref{f:Ng}) shows the results on $N_{\mathrm g}$ from MC simulations using this ``Boost'' method. The multiplicity is plotted as a function of $k_{\perp \mathrm{g}}$, but for each value of $k_{\perp \mathrm{g}}$ results for several values of $E_{\mathrm g}$ are shown. The crosses are mostly on top of each other, confirming that the method gives unbiased gluon jets depending on one scale only.

\begin{figure}[t]
\begin{center}
\epsfxsize=0.7\textwidth
\epsfbox{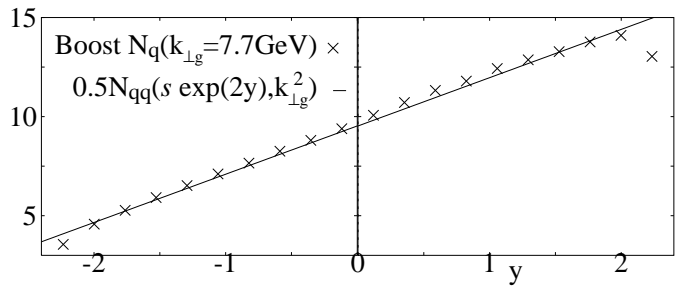}
\end{center}
\caption{MC simulation of quark jets with the Boost definition (crosses). For fixed $k_{\perp \mathrm{g}}$, the $y$ dependence is linear and in agreement with Eq.~(\protect{\ref{e:Nqqbias}}) (solid line). \label{f:Nqy}}
\end{figure} 
The Boost method enables also a study of the $Y_{\mathrm q}$ dependence of the biased multiplicity for fixed $k_{\perp \mathrm{g}}$, complementary to the study suggested in section~\ref{sec:2jet}, where the $k_{\perp r}$ dependence for fixed ``rapidity scale'' $\ln s$ is studied.
Fig.~(\ref{f:Nqy}) shows that MC results for the quark jets in the Boost method agree with the analytical expression in Eq.~(\ref{e:Nqqbias}).
 
\section{Summary}\label{sec:summary}
Event selection introduces bias on multiplicities in two-jet and three-jet events. An expression for this bias,\cite{qgjets} valid in the modified leading log approximation, awaits confrontation with data.
Taking the bias into account, the unbiased multiplicity $N_{\mathrm{gg}}$ can be extracted as $2(N_{\mathrm{q\overline{q}g}}-N_{\mathrm{q\overline q}}(s,k_{\perp \mathrm{g}}^2))$.\cite{letter}

In general, jets are biased, and their properties depend on two scales, transverse momentum and an available rapidity range. 
Unbiased gluon jets where the two scales coincide can, however, be defined by the ``Boost'' boundary definitions.\cite{qgjets}
The Boost method also gives quark jets where both scales, though different, are well defined. An expected linear dependence on the rapidity scale is seen in MC simulations.

\section*{Acknowledgments}
I thank G\"osta Gustafson and  Valery Khoze for a fruitful collaboration.

\end{document}